\documentclass[english]{emulateapj}
\usepackage[T1]{fontenc}
\usepackage[latin1]{inputenc}
\usepackage{graphicx}
\usepackage{amssymb}

\makeatletter

\usepackage{times}
\usepackage{amsmath}

\newcommand{\yr}{{\,\rm yr}}

\newcommand{\pc}{\,\mathrm{pc}}

\newcommand{\AU}{{\,\rm AU}}

\newcommand{\Mbh}{M_{\bullet}}

\newcommand{\Mo}{M_{\odot}}

\newcommand{\rads}{\,{\rm radians}}

\shorttitle{ECCENTRIC DISK INSTABILITY}
\shortauthors{MADIGAN, LEVIN AND HOPMAN}

\usepackage{babel}
\makeatother
\begin{document}
\bibliographystyle{apj.bst} 

\title{A New Secular Instability of Eccentric Stellar Disks Around Supermassive
  Black Holes,\\ 
with Application to the Galactic Center}

\author{Ann-Marie Madigan, Yuri Levin and Clovis Hopman}

\affiliation{Leiden Observatory, Leiden University, P.O. Box 9513, NL-2300 RA Leiden}

\email{madigan@strw.leidenuniv.nl}

\begin{abstract}
We identify a new secular instability of eccentric stellar disks around supermassive black
holes. 
We show that retrograde precession of the stellar orbits, due to the
presence of a stellar cusp, 
induces coherent torques that 
amplify deviations of individual orbital eccentricities from the average, and thus drive all eccentricities away from
their initial value.
We investigate the instability 
using $N$-body simulations, and show that it can propel
individual orbital eccentricities to significantly higher or lower 
values on the order of a precession time-scale.

This physics is relevant for the Galactic center, where massive stars are likely to form in
eccentric disks around the SgrA* black hole. We show that the observed bimodal eccentricity distribution of disk stars in the Galactic center is in good agreement with the distribution resulting from the eccentricity instability and demonstrate how the dynamical evolution of
such a disk results in several of its stars acquiring high ($1-e\ll 0.1$)
orbital eccentricity.
Binary stars on such highly eccentric orbits would get tidally disrupted by the
SgrA* black hole, possibly producing both S-stars near the black hole and
high-velocity
stars in the Galactic halo.
\end{abstract}


%%%%%%%%%%%%%%%%%%%%%%%%%%%%%%%%%%%%%%%%%%%%%%%%%

\section{Introduction}
Most well-observed galaxies with stellar bulges show evidence for supermassive black holes (SMBHs) in their centers \citep*{Kormendy:2004p2865}. An important feature of stellar dynamics in galactic nuclei is that for orbits contained well within the SMBH's radius of influence, the precession time $t_{\rm prec}$ 
is many orders of magnitude larger than their period $P$.\footnote{This is in contrast
to the rapidly precessing stellar orbits typical 
for globular clusters or galaxies.} Therefore,
for a study of long-term (secular) gravitational dynamics, it is useful to think 
of the stars not as particles but as massive elliptical wires stretched along the stars'
eccentric
orbits  \citep{Rauch:1996p270}. Gravitational torques between the
wires can become the dominant mode of 
 orbital eccentricity evolution, as 
is seen in the case of resonant relaxation \citep{Rauch:1996p270,  Gurkan:2007p42, Touma:2009p4317,Eilon:2008p2923}. Secular effects have a significant impact  on the stability of
stellar cusps \citep{Tremaine:2005p2846, Polyachenko:1981p3004,
  Polyachenko:2008p2804},  on gravitational wave emission 
from in-spiraling compact objects \citep*{Hopman:2006p1864, Hopman:2006p1668},
on tidal disruption of stars 
\citep{Rauch:1998p2872} and, as we show here, on the origin of the S-stars in the Galactic center. \\

In this Letter we describe a mechanism for generating high-eccentricity stars
via a fast secular instability of an eccentric 
stellar disk (\S\ref{sec:ins}). We use $N$-body simulations to investigate the
non-linear development of the instability 
(\S\ref{sec:sims} and \S\ref{sec:results}) and to explore its dependence on
the initial eccentricity and 
mass of the disk. We then show that the instability may be relevant
for young stars in the Galactic center,
and may provide an interesting channel for formation of the S-stars and of
high-velocity stars in the Galactic halo  
(\S\ref{sec:sstars}). We critically discuss our results in \S\ref{sec:discuss}.

\section{The eccentricity instability}\label{sec:ins}

Consider a thin stellar disk of mass $M_{\rm disk}$ around a SMBH of mass $\Mbh$ which is embedded in a power-law
stellar cusp of mass $M_{\rm cusp}$. Our two main assumptions are that (1) $M_{\rm disk}\ll M_{\rm
  cusp}$,
so that precession of individual orbits is driven by the cusp, and
(2) initially the stellar eccentricity vectors 
\begin{equation}
{\bf e} = \frac{1}{G \Mbh} {\bf v} \times ({\bf r} \times {\bf v}) - \hat{\bf r}
\end{equation}
are aligned and similar in magnitude so that the disk
precesses coherently as a whole (for a typical power-law cusp, the
orbital precession rate depends only weakly on the semi-major axis; see below).
  Here $|{\bf e}|$ corresponds to
the magnitude of the eccentricity of the orbit and ${\bf r}$ and ${\bf v}$ are the position and
velocity vectors respectively. 
Both assumptions are realistic for 
a disk of young stars formed in a galactic center via the gravitational capture of part of 
an infalling molecular cloud: observations reveal a massive spherical stellar cusp \citep{Genzel:2003p870, Schodel:2007p1797}, while hydrodynamical simulations tentatively show similar initial
orbital eccentricities of young stars \footnote{\citet{Bonnell:2008p3000} make two large simulations, investigating star-formation in the Galactic center. In their Run 1 they find $0.6 < e < 0.7$, while in Run 2 the inner edge of the disk is circularized, and hence the stellar eccentricities range between $0$ in the inner disk and $\sim 0.6$ in the outer. While the precise nature of the circularization, or its absence, has not been investigated, it is clear that it depends on the initial conditions.} \citep{Sanders:1998p2967, Alexander:2008p1407,
  Bonnell:2008p3000, Hobbs:2009p4319}.

The orbits precess in the direction opposite to the orbital rotation of the stars. 
The precession time-scale scales as
\begin{equation} \label{eqn:t_prec}
\begin{split}
t_{\rm prec} & \sim \frac{\Mbh} {N( < a) m} P(a) f(e)\\
& \propto a^{\gamma-3/2} f(e)
\end{split}
\end{equation}
\noindent where $P(a) = 2 \pi \sqrt{a^3/G\Mbh}$ is the period of a star with
semi-major axis 
$a$, $N(<a)$ is the number of stars in the cusp within $a$, $m$ is the individual mass of
the stars, and $\gamma$ 
is the power-law index for the space-density $\rho(r) \propto
r^{-\gamma}$. For a power-law cusp with $\gamma = 3/2$ \citep{Young:1980p4271}, 
$t_{\rm prec}$ is constant for all $a$. 

Observations of the Galactic center indicate that there is a stellar cusp in the inner parsec, with published measurements of the slope not significantly different from $\gamma=3/2$, e.g. $\gamma = 1.4 \pm 0.1$ \citep{Genzel:2003p870} and $\gamma = 1.19 \pm 0.05$  \citep{Schodel:2007p1797}. It is important to stress that these observations only show a small and biased subset of the stellar content in that region. They are restricted to massive young stars, that have not yet relaxed, and to giants, which have relaxed, but are affected by hydrodynamical collisions with other stars \citep[e.g.][]{Ale99}. These two stellar types are not expected to dominate the density. Instead, theoretical models accounting for mass-segregation show that the density at the regions of interest is mostly determined by stellar black holes and main sequence stars, with cusp values between $\gamma_{\rm MS}=1.4$ and $\gamma_{\rm BH}=2.0$ \citep{Ale08}. We conclude that both theory and observations 
yield values of $\gamma$ close to $3/2$ and the dependence of $t_{\rm prec}$ on $a$ is weak. 

A key element of the eccentric instability is that $f(e)$ in
Equation (\ref{eqn:t_prec}) is an increasing 
function of eccentricity \citep[e.g.][]{2008gady.book.....B}, so that an orbit that is
slightly more eccentric than average 
will lag behind in precession. Such a star will feel a
strong, coherent torque from the 
other stars in the disk, in the {\it opposite} direction of its angular
momentum vector. As a result, its angular 
momentum decreases in magnitude, causing its eccentricity to increase even
further. In this way, very high 
eccentricities can potentially be achieved. Conversely, if a stellar orbit is
slightly less 
eccentric than the average, it will have a higher precession rate, 
thus experiencing a torque which acts to decrease its eccentricity further. 

We thus find that the initial 
eccentricity vector distribution is unstable, due to a combination of
retrograde precession and $df/de>0$, and that all eccentricities are driven
away from the 
initial eccentricity of the disk. It can be shown
analytically that the initial growth timescale of the eccentric instability scales
as 
\begin{equation}
\tau_{\rm growth} \sim t_{\rm prec} \left(M_{\rm cusp}\over M_{\rm disk}\right)^{1/2}\left[e\sqrt{1-e^2} \frac{d}{de} \left(\frac{1}{f(e)}\right)\right]^{-1/2}.
\label{tauinst}
\end{equation}

For the Galactic center this amounts to several precession
time-scales.

We now demonstrate the eccentricity instability using $N$-body simulations and
look 
at its impact on the young stars in the Galactic center.

%%%%%%%%%%%%%%%%%%%%%%%%%%%%%%%%%%%%%%%%%%%%%%%%%

\section{$N$-Body Simulations}\label{sec:sims}

We summarize a newly-developed code which we have used in these
simulations. Our integrator is based on a 4th-order Wisdom-Holman algorithm 
\citep{Wisdom:1991p593, Yoshida:1990p2251}, where the forces on a particle are
split into a dominant 
Keplerian force and perturbation forces. We use Kepler's equation to integrate the orbit 
\citep{1992fcm..book.....D}, keeping the position of the central object fixed. As we use adaptive time-stepping to deal with
close encounters, 
the algorithm loses its symplecticity and is not strictly energy conserving. We implement time-symmetry to 
reduce the energy error and the $K_2$ kernel  from \citet{Dehnen:2001p1581}
for gravitational softening. 
The overall fractional energy error for these simulations is typically $\sim 10^{-8}$ or less.
We describe the code in detail in an upcoming paper.

\subsection{Initial Conditions}
We have three main components in our simulations. (1) An eccentric disk  with surface density profile $\Sigma \propto r^{-2}$ \citep[cf.][]{Paumard06}, 
consisting of 100 equal-mass stars, with semi-major axes $0.05
  \pc \!\le\! a \!\le\! 0.5 \pc$. We vary the masses of the stars for different simulations.
We initialize the eccentricity vectors of all stars pointing in roughly the
  same direction, 
though they are scattered in their orbital phase. Initial inclinations are distributed between
  $h/r \sim \pm 0.01$. (2) A SMBH of $4 \times 10^6 M_{\odot}$ \citep{Ghez:2008p3027}, and (3) a smooth stellar cusp with density profile index $\gamma = 1.5$ and a
  mass of $0.5 \times 10^6 M_{\odot}$ within 1 pc  
\citep{Schodel:2007p1797, Genzel:2003p870}.  The stars within 
the disk precess due to the cusp and experience a prograde
  general relativistic precession 
which amounts to apsidal precession of the orbit by an angle
\begin{equation}
\delta \phi_{\rm prec} = \frac{6 \pi G\Mbh}{ a (1 - e^2) c^2}
\end{equation}
per orbit. Precession due to self-gravity of the disk is minimal. \\

To verify our results, we also run simulations with twice the number of half-as-massive stars and various softening parameters. We discuss the use of different values of $\gamma$ 
\citep[e.g.][]{Bah76} in \S\ref{sec:discuss}.

%%%%%%%%%%%%%%%%%%%%%%%%%%%%%%%%%%%%%%%%%%%%%%%%%

\section{Results}\label{sec:results}

\begin{figure} [h]
  \centerline{ \includegraphics[angle = -90, scale=0.36]{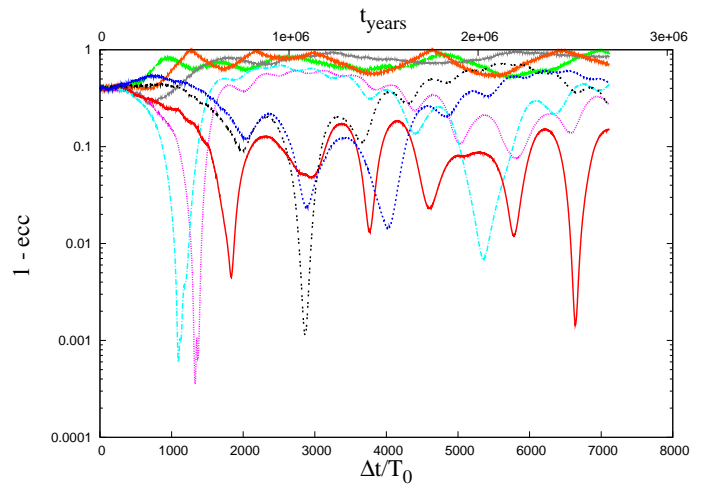}}
 \caption{Eccentricity evolution of stars in an eccentric disk ($M_{\rm disk} = 8 \times 10^3 M_{\odot}$) as a function of time. $\Delta t$ is elapsed time expressed in units of the initial period of the innermost orbit $T_0$ ($t_{\rm prec} \sim 1200 ~ T_0$). The smoothly varying behavior is characteristic of the secular nature of the instability.}
 \label{fig:log_ecc}
\end{figure}

\begin{figure} [h]
  \centerline{ \includegraphics[angle = -90, scale=0.48]{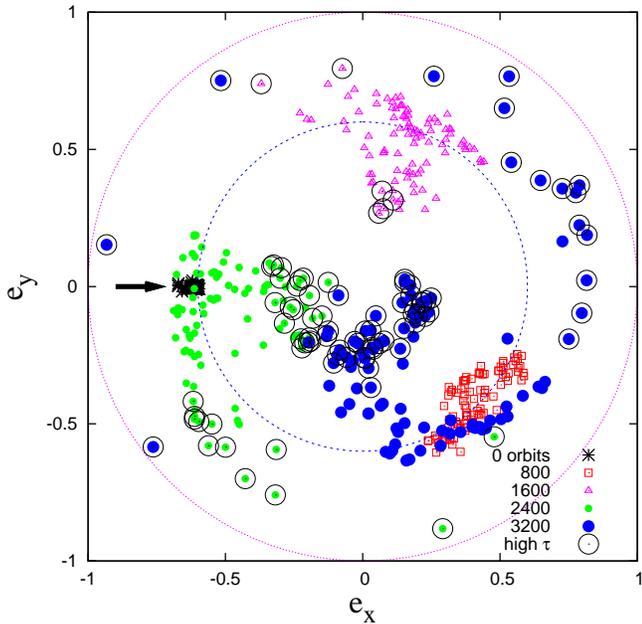}}
 \caption{Evolution of the eccentricity vectors of all the stars in the disk ($M_{\rm disk} = 4 \times 10^3 M_{\odot}$) for different times. The outermost circle indicates where $|{\bf e}| = 1$; the inner circle indicates the initial value of the eccentricities $|{\bf e}| = 0.6$. An arrow points to the initial positions. The eccentricity vectors spread out as the stars complete more and more orbits, and precess with retrograde motion. Those experiencing the highest torques ($\tau$) are over-plotted with small circles. As the eccentric disk instability predicts, the slowest precessing orbits have the highest eccentricities. In addition, orbits with low eccentricity clump together at late times. We believe non-linear effects are responsible for this feature. }
 \label{fig:ev_evo}
\end{figure}

With a mass of $8 \times 10^3 M_{\odot}$ and an initial eccentricity of 0.6,
we let the disk of stars evolve for 
several precession times ($t_{\rm prec} \sim$ 0.6 - 0.7 Myrs or $\sim$ 1200 orbits at $a$ = 0.05 $\pc$). 
In Figure (\ref{fig:log_ecc}) we show a selection of stars from this run which best demonstrates the dispersion of the disk's initial eccentricity, while Figure (\ref{fig:ev_evo}) follows the evolution of the 
eccentricity vectors (this time for $M_{\rm disk} = 4 \times 10^3 M_{\odot}$). The circles in the latter plot indicate the stars
experiencing the highest torque 
($\tau$) and hence angular momentum changes. Note that the highest
eccentricity orbits are precessing 
behind the bulk of the stars as expected. We find a monotonically decreasing relationship between the rate of change of
angular momentum, 
i.e. torque on each orbit, and semi-major axis $a$.

The effects of the instability on the stellar eccentricity distribution are evident after $\sim \! 0.5\!-\!1 ~t_{\rm prec}$ and the strongest torques are experienced over a few precession times before the disk loses coherence. \\

We explore the dependence of the final range of eccentricities on the mass and initial eccentricity of the disk. To this end, we run a set of simulations with a disk of initial eccentricity 0.6 for various disk masses (4, 6, 8, 10 $ \times 10^3 M_{\odot}$), and a set with $M_{\rm disk} = 10^4 M_{\odot}$ for various initial eccentricities (0.1, 0.4, 0.5, 0.6, 0.7, 0.9). The results are presented in Table (\ref{tab:t1}). Higher initial eccentricities and masses of the disk are correlated with higher resulting eccentricities. We also see a correlation between the highest eccentricities and inclinations reached, with the largest values occurring at the smallest radii. \\

In contrast to the secular instability described in
\citet{Tremaine:2005p2846}, relativistic effects in 
this case do not promote stability once the relativistic prograde precession
rate exceeds that of the retrograde precession due to the cusp. The highest eccentricity 
orbit precesses with prograde motion at a higher rate than the average and again 
experiences a torque in the opposite direction of its angular momentum.

%%%%%%%%%%%%%%%%%%%%%%%%%%%%%%%%%%%%%%%%%%%%%%%%%

\section{Application to S-Stars}\label{sec:sstars}

The region surrounding the SMBH in the Galactic center \citep{Gillessen:2009p4320} is host to several interesting
populations of young stars, which include one well-defined clockwise disk of
O and Wolf-Rayet (WR) stars, and a group of counter-clockwise moving O and WR
stars,
possibly a dissolving second disk \citep{Levin:2003p962, Nayakshin:2005p2624, Genzel:2003p870, Paumard06, lu08, Bartko:2008p2762}. These O
and WR stars are located at projected distances of $0.05 \pc$ to $0.5 \pc$, and
were
most likely formed as a result of gravitational fragmentation of a gaseous disk
\citep*{Levin:2003p962}.

\begin{table}[b]
\caption{Simulation Parameters and Statistics of S-stars}
\begin{tabular}{llllllll}
\hline
\hline
$M_{\rm disk}$ &   $e _{\rm disk}$   & $e_{\rm high}$$^a$ & $i_{\rm high}$$^b$ & $a_{\rm bin}$  & $r_{t}$ &  $a_{\bullet}$ & $r_p < r_t$$^c$ \tabularnewline
$[\Mo]$   &  &  & [rads] & [AU]  &  [pc]   & [pc]  &  [$\%$] \tabularnewline
\hline
$4 \times 10^{3}$ 	& 0. 4 & 0.8755 & 0.4662 & 0.1 & $3.57 \times 10^{-5}$ & 0.0026 & 0  	\tabularnewline
	...			& ...	 & ... & ...& 1	& $3.57 \times 10^{-4}$ &  0.026  & 0	\tabularnewline
	...			& 0.6 & 0.9691 & 0.7960 & 0.1 & $3.57 \times 10^{-5}$ & 0.0026 & 0    	\tabularnewline
	...			& ...	 & ...& ...& 1	& $3.57 \times 10^{-4}$ &  0.026  & 0 	\tabularnewline

$6 \times 10^{3}$	& 0.6 & 0.9981 & 1.824 & 0.1 & $3.57 \times 10^{-5}$ & 0.0026 & 4   	 \tabularnewline
   	...			& ...	 & ... & ...& 1	& $3.57 \times 10^{-4}$ &  0.026  & 8	\tabularnewline

$8 \times 10^{3}$ 	& 0. 4 & 0.8796 & 0.6643 & 0.1 & $3.57 \times 10^{-5}$ & 0.0026 & 0 	 \tabularnewline
	...			& ...	 & ... & ... & 1	& $3.57 \times 10^{-4}$ &  0.026  & 0 	\tabularnewline
	...			& 0.6 & 0.9993 & 2.170 & 0.1 & $3.57 \times 10^{-5}$ & 0.0026 & 3    	\tabularnewline
	...			& ...	 & ... & ... & 1	& $3.57 \times 10^{-4}$ &  0.026  & 7	\tabularnewline
	
$1 \times 10^{4}$	& 0.1 & 0.5115 & 0.2935 & 0.1 & $3.57 \times 10^{-5}$ & 0.0026 & 0	\tabularnewline
   	...			& ...	 & ... & ... & 1	& $3.57 \times 10^{-4}$ &  0.026  & 0 	\tabularnewline
	
   	...			& 0.4 & 0.9580 & 1.440 & 0.1 & $3.57 \times 10^{-5}$ & 0.0026 & 1   	\tabularnewline
	...			& ...	 & ... & ... & 1& $3.57 \times 10^{-4}$ &  0.026  & 2 	\tabularnewline
	
	...			& 0.5 & 0.9773 & 1.348 & 0.1 & $3.57 \times 10^{-5}$ & 0.0026 & 0    	\tabularnewline
	...			& ...	 & ... & ... & 1	& $3.57 \times 10^{-4}$ &  0.026  & 3 	\tabularnewline
	
	...			& 0.6 & 0.99991 & 2.328 & 0.1 & $3.57 \times 10^{-5}$ & 0.0026 &14    	\tabularnewline
	...			& ...	 & ... & ... & 1	& $3.57 \times 10^{-4}$ &  0.026  & 14	\tabularnewline
	
	...			& 0.7 & 0.99998 & 3.0312 & 0.1 & $3.57 \times 10^{-5}$ & 0.0026 & 32    	\tabularnewline
	...			& ...	 & ...& ...& 1	& $3.57 \times 10^{-4}$ &  0.026  & 32 	\tabularnewline
	
	...			& 0.9 & 1.0 & 3.125 & 0.1 & $3.57 \times 10^{-5}$ & 0.0026 & 58    	\tabularnewline
	...			& ...	 & ...& ...& 1	& $3.57 \times 10^{-4}$ &  0.026  & 59 	\tabularnewline
\hline
\hline
\multicolumn{8}{l}{$^a$ Mean of highest five eccentricities reached during simulation}\tabularnewline
\multicolumn{8}{l}{$^b$ Mean of highest five inclinations (in radians)}\tabularnewline
\multicolumn{8}{l}{$^c$ Percentage of stars that enter the tidal radius}\tabularnewline 
\multicolumn{8}{l}{
Note that disks with $M_{\rm disk} \gtrsim 6 \times 10^{3} M_{\odot}$ and initial eccentricity }\tabularnewline

\multicolumn{8}{l}{
$e _{\rm disk} \gtrsim 0.6$, produce very high eccentricities, possibly leading to tidal }\tabularnewline
\multicolumn{8}{l}{
disruptions of binaries.
}\tabularnewline
\label{tab:t1}
\end{tabular}
\end{table}

More puzzling in terms of origin is the population of B-stars known as the
`S-stars' \citep{Schodel:2002p3128, Ghez:2005p674, Eisenhauer:2005p3105, Gillessen:2009p4320}. 
At distances of $0.003\!-\! 0.03 \pc$ from the SMBH, in-situ formation seems
highly improbable 
due to the tidal field of SgrA* \citep{Levin:2007p116}. In addition, their young age ($\sim 20 \!-\! 100$ Myr)
imposes a tight
constraint on formation scenarios in that they cannot have travelled far 
from their place of birth. One possibility is the
formation of the S-stars from the disruption of stellar binaries \citep{Hills:1988p3075, Gould:2003p1760}. 
\citet{Perets:2007p3026} propose the surrounding stellar bulge as the source of the binaries, while \citet*{Lockmann:2008p4324} (LBK) invoke
two stellar disks in the 
Galactic center to explain the origin. In their simulations
the stars experienced Kozai-type torques which induced strong changes in their inclination and
eccentricity. The eccentricity of many of the stars became greater than $0.9$,
and LBK argued that if these stars were binaries they would be tidally
disrupted by the SMBH, thus producing the S-stars and high-velocity
stars via Hills' (\citeyear{Hills:1988p3075}) mechanism.
  
However, the gravitational influence of a stellar cusp, which was neglected by LBK,
strongly suppresses Kozai-type dynamics. This was recently discussed in \citet{Chang:2009p4322}. Chang's treatment was highly simplified; in particular, his expressions for Kozai evolution were
valid only if the stellar orbits were inside the inner edges of both
disks. We have performed LBK-type simulations and have confirmed Chang's general conclusion that a
realistically massive stellar cusp entirely suppresses the production of high-eccentricity stars via a Kozai-type
mechanism. \\

In contrast, the eccentricity instability is capable of driving several stars
to near-radial orbits, as is evident in Figure (\ref{fig:log_ecc}). For capture, the pericenter  $r_p$ of the binary's orbit must come within the tidal radius $r_t =(2 \Mbh/m_{\rm bin})^{1/3} a_{\rm bin} $. This results in a captured star with semi-major axis $a_{\bullet}$ that scales as $a_{\bullet} \sim (\Mbh/m_{\rm bin})^{2/3} a_{\rm bin}$, where $a_{\rm bin}$ is the semi-major axis of the binary itself, $m_{\rm bin}$ is its combined mass and the eccentricity of the captured star can be approximated as $1\!-\! e \sim (m_{\rm bin}/\Mbh)^{1/3}$. \\

We summarize our results for varying masses and initial eccentricities of the disk in Table (\ref{tab:t1}). All simulations have evolved for $\sim 7$ Myr. For a given semi-major axis of a binary system, we calculate the tidal radius at which the binary will be disrupted and finally the percentage of stars that come within this radius. It is clear that for many parameters of our disk, stars have $r_p < r_t$. For example, a disk with an initial eccentricity of 0.6 and mass of $10^{4} M_{\odot}$, 14 $\%$ of stars in the disk will enter the tidal radius. Thus, if some of these stars are binaries, Hills' mechanism for producing the S- and high-velocity stars \citep{Brown:2006p4262} could be at work. An important element in this scenario is the binary survival rate at the radii of the disk; however, \citet{Perets:2009p4326} show that for $a = 0.1 \AU$ the evaporation rate is $> 10^7 \yr$. \\

The scalar resonant relaxation time in the Galactic center is everywhere larger than the age of the disks. It follows that since the post-capture eccentricity of the stars is of order $1-e\approx10^{-2}$, most of the S-stars cannot have formed in the {\it current} disk. The lifetime of many S-stars could be of
the order of 100 Myr, and thus earlier ($\sim10\!-\!100$ Myr) starburst episodes could have contributed to their present population. In that case, the scalar relaxation time is short enough to randomize the eccentricities \citep{Hopman:2006p1864, Perets:2009p4325}. We note that the scenario presented by LBK has the same limitation.

\subsection{Bimodal Eccentricity Distribution in Disk}

\begin{figure} [h]
  \centerline{ \includegraphics[clip=true, trim = 6cm 2cm 0 0, angle = -90, scale=0.36]{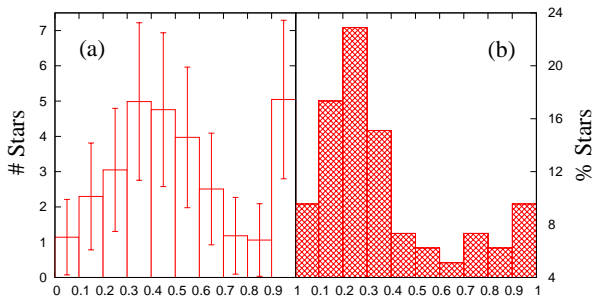}}
 \caption{(a) Observed eccentricity distribution of clock-wise stellar disk in the Galactic center, reproduced with permission from \citet{Bartko:2008p2762}. (b) Simulated distribution for a disk with an initial eccentricity of 0.6 and $M_{\rm disk} = 10^4$  after $5.7$ Myr. Both the observed and simulated eccentricity distributions exhibit bimodality and are in good agreement.}
 \label{fig:bartko_comp}
\end{figure} 

In Figure (\ref{fig:bartko_comp}) we compare the observed stellar eccentricity distribution of the clock-wise disk \citep{Bartko:2008p2762} with the simulated post-instability eccentricity distribution for all stars with an initial disk eccentricity of $0.6$ \citep[c.f.][]{Bonnell:2008p3000}. We draw attention to the qualitative similarities in the two plots, in particular the double-peaked profile. This bimodality is a natural consequence of the eccentric disk instability as all eccentricities are driven away from the mean. Without the instability it is unclear how to generate such a distribution. 

%%%%%%%%%%%%%%%%%%%%%%%%%%%%%%%%%%%%%%%%%%%%%%%%%

\section{Critical discussion}\label{sec:discuss}

So far, we have described the eccentricity instability for an idealized initial disk configuration. We now demonstrate the effectiveness of the instability for a broader scope of parameters and extend our disk model to include (1) a significant range in both magnitude and direction of the stellar eccentricity vectors, (2) heavily inclined stellar orbits, and (3) different values of $\gamma$, the power-law index of the cusp, which determines the precession rate within the disk. We discuss each of these in turn. \\

(1) We simulate disks with initial eccentricity vectors scattered between a range of opening angles $\theta \in [0, 2 \pi ]$. As anticipated, significant spreading of the vector decreases the effectiveness of the instability. However, we find sufficiently high eccentricity orbits to generate S-stars up to $\theta \sim \pi \rads$, which suggests that the instability remains important for substantially less favorable circumstances. Next we vary the magnitudes of the eccentricities throughout the disk, basing our initial conditions on Run 1 and 2 of \citet{Bonnell:2008p3000}. Run 1 (0.6 $<$ $e$ $<$ 0.7) is in excellent agreement with our previous results. The instability in Run 2 (0 $<$ $e$ $<$ 0.6) however, is not efficient at generating the highest $e$ orbits and hence the S-stars.
These conclusions are transparent; the stars which experience the greatest torque are those at small $a$ (see \S\ref{sec:sims}) and if initially circular, need a greater torque to produce the highest $e$ orbits. We find that the most important condition, for $M_{\rm disk} = 10^4$, is for innermost orbits to have $e > 0.4$. Future work on gas dynamics of eccentric disks will tell whether this presents a serious problem for our scenario.\\

(2) We study disks with stellar inclinations $h/r$ in the range $[-0.3, 0.3]$ and find that the instability is robust in all cases. However, above the limit $[-0.2, 0.2]$, the instability appears ineffective in generating the highest $e$ ($> 1 - 10^{-3}$) orbits. \\

(3) We vary the power-law slope of the cusp. For $\gamma < 3/2$, due to the shallowness of the cusp, the innermost orbits lag behind in precession and are more susceptible to being pushed to high $e$ orbits. While this improves the ability of the instability to produce S-stars, the instability is not as effective at generating low $e$ orbits. Conversely, an increase in $\gamma$ $(> 3/2)$ results in the innermost orbits precessing faster than the bulk of the stars and hence suppresses the generation of the highest $e$ orbits and lengthens the time-scale on which it occurs. We conclude that the production of S-stars is unlikely for a cusp with $\gamma > 7/4$. In all cases examined ($1 < \gamma < 2$), the instability proves robust and, as emphasized in \S\ref{sec:ins}, there is both observational and theoretical evidence that the slope of the cusp in the Galactic center is sufficiently close to $3/2$ for the instability to be highly effective. \\

The eccentricity instability described in this work 
is relevant for an eccentric disk embedded in a stellar cusp. We note
that the eccentric disk in M31 \citep*{Tremaine:1995p3101} is not subject to our instability, since
it is not embedded in any visible cusp. Indeed we find that such a disk remains coherent over several precession times, although the innermost stars can undergo significant angular momentum changes as described by \citet{Cuadra:2008p3486}. The situation is different
in the Galactic center, where massive stars were likely born from an
eccentric gaseous disk, and where there is strong observational
evidence for a heavy spherical stellar cusp \citep{Genzel:2003p870, Schodel:2007p1797}. We have shown that the instability 
may have strongly influenced the development of the eccentricities of the massive stars observed, and in particular, can account for the bimodality in their eccentricity distribution. Furthermore, combined with Hills' (\citeyear{Hills:1988p3075}) mechanism, the eccentricity instability may have played a role in the formation of the S-stars and of the high-velocity
 stars in the Galactic halo.

\acknowledgments{We thank Atakan G\"urkan, Richard Alexander, Simon Portegies Zwart, Mher Kazandjian, Jihad Touma and Sergei Nayakshin for
  useful discussions, and Mark Hayden for comments on the text. AM is supported by a TopTalent fellowship from the Netherlands
Organization for Scientific Research (NWO), 
YL by a Vidi fellowship from NWO and CH by a Veni fellowship from
NWO.}

\end{document}